\documentclass[runningheads,a4paper]{llncs}

\usepackage{setspace}
\usepackage{amssymb}
\usepackage{amsmath}
\usepackage{algorithm2e}
\usepackage{nicefrac}

\newcommand{\expected}[1]{\mathbb{E}[#1]}
\newcommand{\var}[1]{\mathrm{Var}[#1]}

\begin{document}

\title{Bounding the Estimation Error of Sampling-based Shapley Value
Approximation}

\titlerunning{Bounding the Estimation Error of Shapley Value Approximation}

\authorrunning{S. Maleki, L. Tran-Thanh, G. Hines, T. Rahwan and A. Rogers}

\author{Sasan Maleki\inst{1} \and Long Tran-Thanh\inst{1} \and Greg Hines\inst{1} \and Talal Rahwan\inst{2} \and Alex Rogers\inst{1}}

\institute{Faculty of Physical Sciences and Engineering \\
University of Southampton, United Kingdom\\
\email{\{sm9g09, ltt08r, gh2, acr\}@ecs.soton.ac.uk}
\and
Masdar Institute of Science \& Technology, United Arab Emirates\\
\email{trahwan@masdar.ac.ae}}


\maketitle

\begin{abstract}
The Shapley value is arguably the most central normative solution concept in
cooperative game theory. It specifies a unique way in which the reward from
cooperation can be ``fairly'' divided among players. While it has a wide range of real
world applications, its use is in many cases hampered by the hardness of its computation.
A number of researchers have tackled this problem by (i) focusing on
classes of games where the Shapley value can be computed efficiently, or (ii)
proposing representation formalisms that facilitate such efficient computation,
or (iii) approximating the Shapley value in certain classes of games.
For the classical \textit{characteristic function} representation,
the only attempt to approximate the Shapley value for the general class of
games is due to Castro \textit{et al.} \cite{castro}. While this algorithm
provides a bound on the approximation error, this bound is \textit{asymptotic}, meaning that it only
holds when the number of samples increases to infinity. On the other hand, when
a finite number of samples is drawn, an unquantifiable error is introduced,
meaning that the bound no longer holds. With this in mind, we provide non-asymptotic bounds on the
estimation error for two cases: where (i) the \textit{variance}, and (ii) the
\textit{range}, of the players' marginal contributions is known.
Furthermore, for the second case, we show that when the range is significantly large
relative to the Shapley value, the bound can be improved (from
$O(\nicefrac{r}{m})$ to $O(\sqrt{\nicefrac{r}{m}})$). Finally, we
propose, and demonstrate the effectiveness of using stratified sampling for
improving the bounds further.
\end{abstract}


\section{Introduction}



The Shapley value is a key solution concept in cooperative game theory.
To date, this value has been applied to a multitude of problems, including
 resolving political conflicts \cite{engelbrecht}, identifying key members in
 terrorists networks \cite{lindelauf,Michalak:etal:13}, analysing customer
 satisfaction \cite{Conklin:2004}, factorizing the risk of diseases
 \cite{Land:2000}, and sharing costs in problems such as railways infrastructure
 \cite{Fragnelli:2000} and multicast transmission \cite{Feigenbaum:2001}. The
 large literature that has grown out of Shapley's paper is a testimony to its
 importance both theoretically and practically.
 
The large literature that has grown out of Shapley's original paper
\cite{shapley_53} is a testimony to its importance both theoretically and
practically. Nevertheless, in the general case, computing the Shapley value has
an exponential time complexity, thus limiting its use to games with relatively
small numbers (e.g., tens) of players. To address this issue, some researchers
have focused on certain, restricted classes of games for which they were able to
develop efficient exact algorithms \cite{ando,Deng:94}.
Other researchers have focused on other restricted classes of games for which
they propose approximate solutions \cite{owen_multilinear,bachrach,nowell}.
Another related line of research has focused on computing the Shapley value using
alternative representation formalisms, as opposed to the standard characteristic
function representation where no assumption is made about how coalition values
are computed \cite{Ieong:Shoham:05,Elkind:2008,Aadithya:11}.

As for the general class of games in characteristic function form, the only
approximation algorithm in the literature is due to Castro \textit{et al.}
\cite{castro}. In particular, they use random sampling to approximate the
Shapley value in games where the variance of the marginal contributions of the
players is known. At first glance the bound that they establish, based on the
central limit theorem (CLT), may seem correct. However, we highlight in this
paper an error term that is not factored in by Castro \textit{et al.} in their
bound. This error is due to the asymptotic assumption that lies at the heart of
the CLT; the results therein only hold as the sample size \textit{increases to
infinity}. Conversely, when the CLT is applied with a finite number of samples
drawn from an unknown distribution, it is known that an unquantifiable
error to the bound will be introduced (see Section \ref{section:castro_clt} for
a detailed explanation). With an unquantifiable error, the possibility of a
significant impact on the bound cannot be ruled out. As such, ignoring this
error results in an inaccurate bound. 

In summary, despite the growing research on the computational aspects of game
theory, and in particular cooperative game theory \cite{george_edith_mike}, the
problem of approximating the Shapley value for the general class of games is
still open. By ``the general class of games'' we shall mean the class of
transferable utility cooperative games in characteristic function form where the
value of a coalition is given by an oracle.

Ideally, an approximation algorithm would establish a tight bound on its error.
The quality of this bound depends on how much information is available from the
problem. Let us observe that in the general class of games, there is no
information that can be exploited to establish a bound on the Shapley value.
Therefore, we have to assume that at least some limited information is available
from the problem instance.

Against this background, we use existing results from probability theory to
approximate and establish a bound on the Shapley value of a player.
Specifically, we use two techniques, namely (i) simple random sampling, and (ii)
stratified sampling, to estimate the Shapley value. In the former, samples are
drawn uniformly from a population, while in the latter, a population is
stratified (i.e., partitioned into ``strata''), such that samples are drawn from
each stratum independently. Furthermore, we use non-asymptotic concentration
inequalities to bound the estimation error. Specifically, our contribution to
the state of the art can be summarized as follows:
\begin{itemize}
  \item In the context of simple random sampling, we consider two
  cases, depending on the information available regarding the marginal
contributions of a player: (i) when their variance is known, we use
Chebyshev's inequality; and (ii) when their range (i.e., the difference
between the minimum and maximum marginal contributions) is known, we use Hoeffding's inequality
\cite{hoeffding}. 
As opposed to Castro
\textit{et al.}'s bound, which only holds asymptotically (i.e., when the number
of samples increases to infinity), our bounds hold when the number of
samples we draw is finite.

\item We show that when the range of the marginal contributions is known, and is
significantly large relative to the Shapley value, the bound can
be improved from $O(\nicefrac{r}{m})$ to
$O(\sqrt{\nicefrac{r}{m}})$, where $r$ is the range and $m$ is the
sample size.

  \item We propose the use of stratified sampling for approximating the Shapley
  value. As a proof of concept, we demonstrate the effectiveness of this
  technique, by applying it to a class of games where every coalition's value is
  bounded by a linear function of the coalition's size.
\end{itemize}

\noindent Note that the use of Hoeffding's and Chebyshev's inequalities were also
proposed by others in the literature (see Section~\ref{section:related_work}). However,
these were only used for simple games and supermodular games. Also note that
our sampling algorithm for the first two cases is similar to Castro \textit{et
al.}'s \cite{castro}; the difference lies in the bounds that we provide. As
such, the novelty of our paper lies in the introduction of the said inequalities
to approximating the general case of the Shapley value, as well as the use of
stratified sampling to obtain a more efficient bound.

The rest of the paper is organised as follows.
Section~\ref{section:preliminaries} provides preliminaries and basic
notations. Section~\ref{section:related_work} gives an overview of the related
work. Section~\ref{section:simple_random_sampling} discusses how the estimation
error can be bounded using simple random sampling.
Section~\ref{Section:stratified} demonstrates the potential of using
stratified sampling to approximate the Shapley value. Finally,
Section~\ref{section:conclusions} concludes the paper and outlines the future
direction.
\section{Preliminaries}
\label{section:preliminaries}
Given a set, $N$, of $n$ players, each labeled by $i$ ($i \in \mathbb{N}$), a
\textit{coalition} $C$ is defined as a subset of $N$. The coalition
$N$ is called the \textit{grand coalition}. Any coalition has a value which is
given by a \textit{characteristic function} $v$. This function maps each subset
of $N$ to a real number, i.e., $v : 2 ^ {N} \rightarrow \mathbb{R}$. The pair
$(N, v)$ specifies a unique cooperative game in characteristic function form.

Given a player, $i$, and a coalition, $C\subseteq N\backslash\{i\}$, the
\textit{marginal contribution} of $i$ to $C$, is the difference in value that
$i$ makes by joining $C$, which is calculated as: $v(C\cup \{i\})
- v(C)$.

A coalition consisting of $n$ players can form in $n!$ ways if we consider all
the possible joining orders of the players. In each joining order, as a player
steps in the coalition, it makes a marginal contribution to the players
that have joined before him. Viewing this joining process as a stochastic
variable, with all joining orders having the same probability of forming (i.e., $1/n!$), the Shapley value
of a player would be his expected marginal contribution. Denote by $\pi(N)$ the
set of all permutations of the players in $N$, each of which representing a distinct joining order. Furthermore, denote by
$P_i^O$ the set of players that precede $i$ in the permutation $O$. The Shapley
value of $i$ can be calculated as:
\begin{eqnarray}
\nonumber
\phi [i, v] & = & \frac{1}{n!} \displaystyle \sum_{O \in \pi(N)}\left [
v(P_i^O \cup \left \{ i \right \}) - v(P_i^O) \right ]\\
\label{eq:shapley}
& = & \displaystyle \sum_{C \subseteq N \setminus \{i\}} \frac{|C|!\;
(n-|C|-1)!}{n!}(v(C\cup\{i\})-v(C)) \\
\label{eq:shapley_v2}
& = & \displaystyle \frac{1}{n}\sum_{k=0}^{n-1}{\sum_{|C| =k, \{i\} \not\in C}{\frac{1}{{n-1\choose k}}(v(C\cup\{i\})-v(C)) }}
\end{eqnarray}

The Shapley value satisfies the axioms of \textit{symmetry}, \textit{efficiency}
and \textit{additivity}. Symmetry states that if two players make the same
marginal contributions to any coalition, their value is equal. The efficiency
axiom implies that the value of the grand coalition is fully divided.
Finally, the additivity means that if a new game is obtained by adding two
different games with the same set of players, the value of a player in the new
game is equal to the sum of his values in the two games. These three axioms
define the `value' of a game uniquely \cite{shapley_53}.

\section{Related Work}
\label{section:related_work}
 
As we mentioned in the introduction, almost all of the existing research on
approximating the Shapley value focused on restricted classes of games. In this
section, we give a brief overview of the notable works in this direction. We
then proceed by explaining why the CLT cannot be relied upon to establish a
bound on the approximation in the general case as Castro \textit{et al.} do.
Finally, we close this section by giving a background on the concentration
bounds that we will use in our results.

\subsection{Efficiently computing or approximating the Shapley value}

One line of research has focused on studying special classes of games where the
Shapley value can be computed efficiently. For instance, Ando focused on minimum cost spanning tree games \cite{ando}, while Deng and
Papadimitriou focused on induced subgraph games \cite{Deng:94}.

In another line of research, approximation methods have been proposed for
certain classes of games, where the exact computation is intractable. For
instance, Mann and Shapley proposed a Monte Carlo simulation method for
estimating the Shapley value in voting games in linear time
\cite{monte_carlo_shapley}. Owen proposed a multilinear extension method with
which the Shapley value in weighted voting games can be approximated in linear
time \cite{owen_multilinear}. Bachrach \textit{et al.} proposed a sampling
algorithm for approximating the Shapley-Shubik and Banzhaf power indices, and
used Hoeffding's inequality \cite{hoeffding} to establish a bound on their
approximation \cite{bachrach}. Fatima \textit{et al.} put forward a linear-time
approximation algorithm for $k$-majority games \cite{fatima}. Liben-Nowell
\textit{et al.} proposed a polynomial time algorithm for supermodular games, and
used Chebyshev's inequality to bound the approximated Shapley value
\cite{nowell}.

Another relevant line of research is concerned with proposing alternative
representation formalisms, where non-restrictive assumptions are made on the way
coalition values are computed (as opposed to the classical characteristic
function representation where no such assumptions are made).
For instance, the marginal contribution nets representation, proposed by Ieong
and Shoham \cite{Ieong:Shoham:05} and extended by Elkind \textit{et al.}
\cite{Elkind:2008}, allows for computing the Shapley value in time linear in the
size of the input. Similarly, Aadithya \textit{et al.} proposed another
representation based on algebraic decision diagrams, where the Shapley value can
be computed in polynomial time \cite{Aadithya:11}.

\subsection{Estimating the Shapley value using the Central Limit Theorem}
\label{section:castro_clt}
Estimating a population parameter such as the mean is a very common task in
statistics. The typical technique to estimate the population mean is to draw
random samples from the population (known as simple random sampling) and
calculate their mean. If a sufficiently large number of samples is drawn, the
mean of the samples (known as the sample mean) tends to be a good estimate of
the population mean. The typical question in this case is how far the sample
mean can be from the population mean. Often researchers resort to constructing a
\textit{confidence interval} to provide an answer.

If we view the Shapley value as the mean of a population of marginal
contributions of a player, which has an unknown distribution, the link between
the random sampling method and estimating the Shapley value becomes clear.
Castro \textit{et al.} use a simple random sampling algorithm and provide a
confidence interval based on the CLT \cite{castro}. In what follows, we show why
using the CLT results in an inaccurate confidence interval.

If we repeat the process of sampling, each time a possibly different mean is
obtained. This observation allows us to view the sample mean itself as a random
variable with some probability distribution. Informally, the CLT states that as
the number of samples that we draw increases to infinity, the distribution of
the sample mean converges to a normal distribution whose mean is equal to the
mean of the original population, and whose variance is equal to the variance of
the samples divided by the sample size. This raises two crucial questions.
First, what is the convergence rate? Second, if we draw a finite number of
samples, what will be the error, i.e., the difference between the true
distribution of the sample mean and the asymptotic normal distribution?

In theory, one way to obtain the convergence rate and an upper bound on the
error is to use the Berry-Esseen theorem \cite{berry_esseen_theorem}, which,
unfortunately, only holds under some strong assumptions. Unless we know the
population distribution, neither of the above questions can be answered in
practice. Therefore, it is not possible to quantify the error, and as such, ignoring it results in an inaccurate
bound. Note that the closer the population distribution is to normal, the
lower the error would be. However, in the general class of games, the marginal contributions of a player could
follow any distribution.

\subsection{Concentration inequalities}
Concentration inequalities generally associate a probability to a bound on the
expected value of a variable. Several such inequalities have been derived from
Markov's inequality: $\Pr(X \geq a) \leq \expected{X}/a$. For instance, if one
applies Markov's inequality to the random variable $(X-\expected{X})^2$,
Chebyshev's inequality will be obtained, which is as follows:
\begin{equation}
\label{eq:chebyshev}
\Pr(|\overline X - \expected{X}| \geq k \sqrt{\var{X}}) \leq \frac{1}{k^2}
\end{equation}
where $k$ is a constant. Chebyshev's inequality is useful in bounding the
mean of a variable when the only information available about it is its variance.

Another useful inequality is that of Hoeffding, which provides an exponential
deviation bound. Hoeffding's Theorem~2 \cite{hoeffding} states that if $S$ is
the sum of $m$ independent random variables, $X_1, \ldots, X_m$, each of which
is bounded by two values, $a_i$ and $b_i$, then the following inequality holds
about $S$:
\begin{equation}
\label{eq:hoeffding}
\Pr(|S - \mathbb{E}[S]| \geq t) \leq 2 \exp \left( -
\frac{2t^2}{\sum_{i=1}^m (b_i - a_i)^2} \right),
\end{equation}
Hoeffding's inequality is particularly interesting as it implies that the
probability of a large deviation from the mean is exponentially small.

Recently, Vu presented an inequality which is concerned with variables that have
a distribution with have heavy tails \cite{vu}. In particular, it provides a
deviation bound of a function of independent variables, which has a general form
and satisfies a number of conditions (for  more detail, see \cite{vu}). In
Section \ref{section:simple_random_sampling}, we will use a special case of
Vu's general bound, tailored to our problem.

Obviously, since such inequalities can be applied to a random variable with any
arbitrary distribution, a bound developed based on them cannot be matched in
quality to one that is developed based on a given distribution.
\section{Simple Random Sampling}
\label{section:simple_random_sampling}
In this section, we show the estimation error can be bounded in the cases where
the range or the variance of the marginal contributions of a player is given.
Our objective is to estimate the Shapley value of a player with probability at
least $(1 - \delta)$ that our estimation error is $\epsilon > 0$. Hereafter,
unless otherwise indicated, the player index $i$ and the characteristic
function $v$ are omitted from all notations of the Shapley value for the sake of simplicity.
Formally, we are interested in finding the condition for the following
inequality to hold:
\begin{equation}
\label{eq:pac_inequality}
\Pr(|\hat{\phi} - \phi| \geq \epsilon ) \leq \delta
\end{equation}
where $\hat{\phi}$ is the estimated Shapley value of a player of interest.

Using simple random sampling, we estimate $\hat{\phi}$ as the mean of $m$ random
samples, $\phi_1, \dots, \phi_m$, taken from the population of marginal contributions of the
player. That is, $\hat{\phi} = 1/m \sum_{j=1}^m \phi_j$. Based on this,
for inequality~\eqref{eq:pac_inequality} to hold, the condition is that $m$ must be sufficiently large.
We will determine the minimum required value for $m$ in the case where either the variance is known (Section \ref{section:case_variance}) or the range is known (Section~\ref{section:case_range}) for the players' marginal contributions.

The sampling algorithm for the aforementioned cases is presented in
Algorithm~\ref{algo:sampling}. Note that it is possible to re-use a sample
(i.e., permutation) for all players; this will not affect the bound. In doing
so, the number of samples required to approximate the Shapley value for all
players would be equal to the largest number of samples required for any one of those
players.

\begin{algorithm}[tb]
\label{algo:sampling}
\SetAlgoLined
\LinesNumbered
\caption{Simple Random Sampling Algorithm}
\KwIn {Game $(N,v)$, $\epsilon$, $\delta$}
\KwOut {$\hat{\phi}$  \,\, // an array containing the approximated Shapley
values}

$M \gets$ GetLargestSampleSizeAmongAllPlayers($\epsilon$, $\delta$)\\
$\forall i \in N$ ; $\hat{\phi[i]} \gets 0$\\
\For{$sample = 1$ to $M$} {
$O \gets$ GenerateUniformRandomPermutation(N)\\
\ForAll {$i \in N$} {
 $P_i^O \gets$ CalculateSetOfPrecedingPlayers($O$, $i$)\\
 $\hat{\phi[i]} \gets \hat{\phi[i]} + \frac{1}{M} \times (v(P_i^O \cup \left \{
 i \right \}) - v(P_i^O))$  \\
}
}
\end{algorithm}

\subsection{Bounding the estimation error when the variance of marginal contributions is known}
\label{section:case_variance}
Let us observe that the variance of the sum of a number of random variables is equal to the sum of the
variances of individual variables, i.e., $\var{\phi_1+\ldots+\phi_m}
= \var{\phi_1} + \ldots + \var{\phi_m}$. In order to determine the
minimum required sample size when the maximum variance of the marginal contributions of
a player, $\sigma^2$, is given, we can use Chebyshev's inequality as follows.
Let $Y=\sum_{j=1}^m \phi_j$. Since $\hat{\phi}=1/m\times Y$, application
of inequality~\eqref{eq:chebyshev} to $Y$ results in: $\Pr(|Y - \expected{Y}|
\geq k \sqrt{\var{Y}}) =  \Pr(|\frac{1}{m}Y - \phi| \geq \frac{1}{m}k
\sqrt{\var{Y}})$.
Let $\epsilon$ be $(k/m)\sqrt{\var{Y}}$. We have:
\begin{equation*}
 \Pr(|\hat{\phi} - \phi| \geq \epsilon) \leq 
\frac{\var{Y}}{(m\epsilon)^2} =  \frac{\var{\phi_1} + \ldots +
\var{\phi_m}}{(m\epsilon)^2}
= \frac{\sigma^2}{m\epsilon^2}
\end{equation*}
Since our aim is for the right hand side of the above inequality to be at most
$\delta$, it follows that $m \geq \lceil \sigma^2/(\delta\epsilon^2) \rceil$.

\subsection{Bounding the estimation error when the range of marginal contributions is known}
\label{section:case_range}
Given the range, $r$, of the marginal contributions of a player, we can use
Hoeffding's inequality to determine the minimum value of $m$ as follows. In
inequality~\eqref{eq:hoeffding}, let $S =\sum_{j=1}^m \phi_j$. We have:
$\Pr(|S - m\phi| \geq t) = \Pr(|\hat{\phi} - \phi| \geq \frac{t}{m})$. Therefore:
\begin{equation*}
\Pr(|\hat{\phi} - \phi| \geq \epsilon) \leq 2 \exp \left( -
\frac{2m^2\epsilon^2}{mr^2}\right)
\end{equation*}
Since we want the right hand side to be bounded by $\delta$, we have:
\begin{equation*}
2 \exp \left( - \frac{2m\epsilon^2}{r^2}\right) \leq \delta \Rightarrow
\frac{-2m\epsilon^2}{r^2} \leq \ln (\frac{\delta}{2})
\end{equation*}
Therefore, the minimum required sample size is: $m \geq \lceil \frac{\ln
(\frac{2}{\delta})r^2}{2\epsilon^2} \rceil$.

Observe that this estimation error has an upper bound of $O(\nicefrac{r}{m})$. As such, if $r$ is large, it becomes dominant within the bound. Therefore, in this
case, a bound that is sublinear in $r$ would be more preferable. Given this, in
the rest of this section, we aim to provide an improved bound for the estimation
error of the Shapley value.
In particular, we provide a $O(\sqrt{\nicefrac{r}{m}})$ bound, based on
the following theorem.
\begin{theorem}
Suppose that $\phi = O\left(rf(n)\ln{n} \right)$ where $n$ is the number of
players, $0< f(n) < 1$ is an arbitrary function of n, and r is the range of
marginal contributions of a player of interest.
There exist constants $c, d > 0$ such that
$$P( | \hat{\phi} -\phi | \geq \epsilon ) \leq de^{\frac{-c m \epsilon^2}{\phi}}$$ 
\end{theorem}
Note that the condition $\phi = O\left(rf(n)\ln{n} \right)$ reflects the fact
that the marginal contributions might be significantly much larger than the
Shapley value (as with heavy-tailed distributions).
The proof of this theorem is a straightforward application of Corollary 4.7 from ~\cite{vu}, tailored to our problem, and thus, is omitted.
From this theorem, we can show that if $m$ the number of samples our algorithm takes, then with at least $(1-\delta)$ probability, the estimation error of the Shapley value is at most
$$
| \hat{\phi} -\phi | \leq \sqrt{\frac{ \phi \left(-\ln{\frac{\delta}{d}}\right)}{cm}} = O\left( \sqrt{\frac{ r\ln{n}(-\ln{\delta})}{\, m}} \right)
$$
This implies that the estimation error has a form of $O(\sqrt{\nicefrac{r}{m}})$ with high probability.
Although this bound is more efficient in terms of $r$, it is less efficient in
terms of $\frac{1}{m}$. This is the trade-off compared to the previous case.
Note that, in contrast to the results provided in the previous sections, this result does not provide an exact bound, as the value of the coefficients is hidden in the big-O notation.
However, it still gives us a useful insight that even within cases of heavy tail distributions, our algorithm can still provide low (i.e., $O(\sqrt{r})$) estimation error.

\section{Stratified Sampling for Coalitions}
\label{Section:stratified}
In the sampling algorithm from the previous section, we were indifferent about
which area of the population to sample from. However, by dividing the
population into a number of homogeneous subpopulations (i.e., strata), it is possible to optimise the number of samples taken from each strata. Provided that a population has been stratified so that members of any stratum are as close as possible,
using stratified sampling will always achieve higher precision than simple
random sampling. Of course, this gain is higher when differences between the
strata are greater.

Suppose that the population of marginal contributions of a player can be stratified into a number of strata, each containing
marginal contributions with similar values (i.e., values that are close to each other). Our hypothesis is that it is possible to bound the
estimation error more efficiently than the bounds in the previous section.
In so doing, we can use Hoeffding's inequality to efficiently bound the estimation error within each
stratum. Note that within this section, as opposed to the population of permutation of players, we sample from the population of coalitions.

To demonstrate the aforementioned hypothesis, we  
prove that there exist two values, $a$ and $b$, such that
$\forall C \subseteq N$, we have:
\begin{equation} 
\label{eq:bounds}
a|C| \leq v(C) \leq b|C|,
\end{equation}
where $|C|$ is the size of coalition $C$.
In particular, we show that by setting
\begin{eqnarray}
\nonumber
a &=& \min_{C \subseteq N}{\frac{v(C)}{|C|}} \\
\nonumber
b &=& \max_{C \subseteq N}{\frac{v(C)}{|C|}}
\end{eqnarray}
we obtain the requested linear bounds given in~\eqref{eq:bounds}.
To prove this, first observe that $a$ and $b$ exist, due to the fact that $N$ is finite.
By definition, for any $ C \subseteq N$, we have:
\begin{equation}
\nonumber
a \leq \frac{v(C)}{|C|} \leq b
\end{equation}
This implies that the following holds
\begin{equation}
\nonumber
a|C| \leq v(C) \leq b|C|
\end{equation}

Hereafter, we assume that~\eqref{eq:bounds} holds.
Our objective is to distribute $m$ samples among $k$ strata, such that the
total estimation error of the Shapley value is minimised.

Now, without loss of generality, we estimate the Shapley value of player $i$.
By grouping the coalitions that do not contain $\{i\}$, based on their sizes, we have $n$ strata $S_0,
S_1, \dots, S_{n-1}$ such that $S_k = \{C \subseteq N\setminus \{i\}, \left|C\right| = k \}$
contains all the coalitions with size $k$. 
Let $r_{k}$ denote the range of marginal contributions of player $i$ within
group $S_{k}$.
It is easy to show that 
\begin{equation} 
\label{eq:range_bound}
 r_{k} \leq 2(b - a)(|C| + 1) = d\left(k + 1\right),
\end{equation}
where $d = 2(b - a)$ (note that $|C| = k$ as $C \in S_k$).
In particular, let $r_{\mathrm{min}}$ and $r_{\mathrm{max}}$ denote the minimum and maximum marginal contribution value of agent $i$ within group $S_k$.
It is easy to show that
\begin{eqnarray}
\nonumber
r_{\mathrm{max}} &\leq& (b-a)k +b \\
\nonumber
(a-b)k +a &\leq& r_{\mathrm{min}}
\end{eqnarray}
In fact, the first inequality holds because $v(C\cup\{i\}) \leq b(k+1)$ and $v(C) \geq ak$.
Similarly, the second inequality holds because $v(C\cup\{i\}) \geq a(k+1)$ and $v(C) \leq bk$. 
Since $r _{k}= r_{\mathrm{max}} - r_{\mathrm{min}}$, we obtain that $ r_{k} \leq 2(b - a)(k + 1)$.

Let $\phi^k$ denote the expected marginal contribution of the player
within stratum $S_k$. 
That is, we have:
\begin{eqnarray}
\nonumber
\phi^k &=& \frac{1}{|S_k|}\sum_{|C| =k, \{i\} \not\in C}{(v(C\cup\{i\})-v(C))} \\
\label{eq:aux_1}
&=& \frac{1}{{n-1\choose k}}\sum_{|C| =k, \{i\} \not\in C}{(v(C\cup\{i\})-v(C))}
\end{eqnarray}
It is obvious that the Shapley value of this player can be
calculated as follows:
\begin{equation} 
\phi = \frac{1}{n}\sum_{k=0}^{n-1}{ \phi^k }.
\end{equation}
In fact, 
we have:
\begin{eqnarray} 
\nonumber
\phi = \frac{1}{n}\sum_{k=0}^{n-1}{ \phi^k } = \frac{1}{n}\sum_{k=0}^{n-1}{  \frac{1}{{n-1\choose k}}\sum_{|C| =k, \{i\} \not\in C}{(v(C\cup\{i\})-v(C))} }
\end{eqnarray}
which is equivalent to the definition of the Shapley value given in \eqref{eq:shapley_v2}.

Let $\phi$ and
$\phi^{k}$ denote the (global) Shapley value and the expected value of
stratum $k$, respectively. Furthermore, denote by $m_{k}$ the number of samples we take
from stratum $S_k$, and denote by $\hat{\phi}^k_{m_k}$ the estimation of $\phi^{k}$
using $m_k$ samples. Based on Hoeffding's inequality (\eqref{eq:hoeffding}),
we have:
\begin{equation} 
\label{eq:estimation_per_group}
| \hat{\phi}^k_{m_k} - \phi^{k} | \leq \sqrt{\frac{r^2_k\left(-\ln{\frac{\delta}{2}}\right)}{2m_k}} \leq d(k+1)\sqrt{\frac{\left(-\ln{\frac{\delta}{2}}\right)}{2m_k}}
\end{equation}
with at least $(1-\delta)$ probability.  
The second inequality comes from equation~\eqref{eq:range_bound}.
Let $\hat{\phi}$ denote the estimated (global) Shapley value of player $i$, which can be calculated as:
\begin{equation} 
\hat{\phi} = \frac{1}{n}\sum_{k=0}^{n-1}{ \hat{\phi}^{k}}.
\end{equation}
Given this, the total estimation error of $\phi$ can be estimated as follows:
\begin{equation} 
| \hat{\phi} - \phi | \leq \frac{1}{n}\sum_{k=0}^{n-1}{ | \hat{\phi}^{k} - \phi^k |}.
\end{equation}
Equation~\eqref{eq:estimation_per_group} implies that for any $0 < \beta < 1$, by setting $\delta = 1- (1-\beta)^{1/n}$, then with at least $(1-\beta)$ probability, we have:
\begin{equation} 
| \hat{\phi} - \phi | \leq \frac{1}{n}\sum_{k=0}^{n-1}{ d(k+1)\sqrt{\frac{\left(-\ln{\frac{\delta}{2}}\right)}{2m_k}}} = \frac{d\sqrt{\frac{\left(-\ln{\frac{1- (1-\beta)^{1/n}}{2}}\right)}{2}}}{n}\sum_{k=0}^{n-1}{ \frac{(k+1)}{\sqrt{m_k}}}
\end{equation}
Hence, in order to achieve efficiently low estimation error, we aim to minimise the following optimisation problem:

\noindent
Determine $m_1,\dots,m_n$ such that  
$$\min{\sum_{k=0}^{n-1}{ \frac{ (k+1)}{\sqrt{m_k}}}}$$ 
is achieved, subject to 
$$\sum_{k=0}^{n-1}{m_k} \leq m$$ 

Now, note that this is a hard optimisation problem, as $m_k$ has to be integer. However, if we relax the problem such that $m_k$ are allowed to be fractional, then the optimal (fractional) $m_{k}^{*}$ can be easily calculated (e.g., by using the Lagrange relaxation approach).
In particular, we can easily show that $\forall k$:
\begin{equation}
\label{eq:optimal_fractional}
m_k^{*} = \frac{m \left(k+1\right)^{\frac{2}{3}}}{\sum_{j=0}^{n-1}{\left(j+1\right)^{\frac{2}{3}} }} \
\end{equation}

\begin{algorithm}[tb]
\label{algo:strat_sampling}
\SetAlgoLined
\LinesNumbered
\caption{Stratified Random Sampling Algorithm}
\KwIn {Game $(N, v)$, $m$, $\delta$}
\KwOut {$\hat{\phi}$  \,\, // an array containing the approximated Shapley
values}
$\forall i \in N$ ; $\hat{\phi[i]} \gets 0$\\

\ForAll {$i \in N$} {
	$\forall k \in \{1,2,\dots,i-1,i+1,\dots n\}$ ; $S_{k} \gets $ GetAllCoalitionsWithSize($k$) \\
	$\forall k \in \{1,2,\dots,i-1,i+1,\dots n\}$ ;  $m_{k} = \left\lfloor  \frac{m \left(k+1\right)^{\frac{2}{3}}}{\sum_{j=0}^{n-1}{\left(j+1\right)^{\frac{2}{3}} }} \right\rfloor$ \\
	\If {$m - \sum{m_k}  >  0$} {
	       UniformlyDistributeRemainingSamplesAmongStrata() \\ 
	}
	$\forall k \in \{1,2,\dots,i-1,i+1,\dots n\}$ ; $\hat{\phi}^{k}[i] \gets $ AverageOfStratum($k$) \\
	$\hat{\phi}[i] \gets AverageOfAllStrata()$
  }
\end{algorithm}

Given this, we can set the value of $m_{k} = \left\lfloor m_{k}^{*} \right\rfloor$.
However, this implies that there may be additional samples left unused as $\sum_{k=0}^{n-1}{m_k}$ may be lower than $m$. In this case, we sequentially increase the value of $m_k$ with $1$ from $k=0$ to $(n-1)$ until we exceed $m$.
The pseudocode of this stratified sampling algorithm can be found in Algorithm~\ref{algo:strat_sampling}.
Note that it is easy to show that $m_{k} \geq \frac{m_k^{*}}{2}$.
These results together imply that with at least $(1-\delta)^n$, the total estimation error of the Shapley value is at most:
\begin{equation} 
| \hat{\phi} - \phi | \leq \frac{d\sqrt{-\ln{\frac{\delta}{2}}}}{n\sqrt{m}} \left( \sum_{k=0}^{n-1}{ \left( k+1 \right)^{\frac{2}{3}} } \right)^{\frac{3}{2}}.
\end{equation}
By using the generalised inequality between two power means for $\frac{2}{3}$ and $1$, we can further bound this estimation error with:
\begin{equation} 
\label{eq:final_error_bound}
| \hat{\phi} - \phi | \leq \frac{d\sqrt{-\ln{\frac{\delta}{2}}}}{n\sqrt{m}} \sum_{k=0}^{n-1}{ (k+1) } = \frac{d\sqrt{-\ln{\frac{1- (1-\beta)^{1/n}}{2}}}}{\sqrt{m}} \frac{n+1}{2}.
\end{equation}
since $\sum_{k=0}^{n-1}{ (k+1) } = \frac{n(n+1)}{2}$. 
This estimation bound is indeed very efficient, especially when the number of samples $m$ is sufficiently large, compared to  $n$, the number of players.
To demonstrate this, we compare this error bound with the error estimation that we could get by using simple random sampling, but only by using Hoeffding's inequality on the whole $m$  samples. 
It is easy to show that the latter estimation is at least $d\sqrt{n \, (-\ln{\frac{\delta}{2}}) }$, which is significantly higher (i.e., worse), compared to the error bound given in equation~\eqref{eq:final_error_bound}, if $m > \frac{(n+1)^2}{4}$.

\section{Conclusions and Future Work}
In this paper, we considered the problem of approximating the Shapley value in the general class of characteristic function games. While the state-of-the-art algorithm in the literature provides an asymptotic bound, we used two concentration inequalities to establish non-asymptotic bounds. This was done for cases where the variance or the range of the players' marginal contributions is known. We also showed how to improve the bound in cases where the range is significantly large relative to the Shapley value. Finally, we proposed the use of the stratified sampling technique, and demonstrated its effectiveness in improving the bounds. In the future, we will focus on extending the use of stratified sampling techniques to classes of interest such as weighted voting games.
\label{section:conclusions}

{\fontsize{10}{10}\selectfont{
\bibliographystyle{plain}
\bibliography{references}
}}

\end{document}